\documentclass[a4paper]{article}

\usepackage{INTERSPEECH2018}
\usepackage{amsmath,epsfig,balance}
\usepackage{multirow,color}
\usepackage{epstopdf}
\usepackage{subfigure,cite}
\usepackage{stfloats}

\def \etal {et al.}
\def \eg {e.g.,~}

\def \ie {i.e.,~}

\title{Ladder Networks for Emotion Recognition: Using Unsupervised Auxiliary Tasks to Improve Predictions of Emotional Attributes \vspace{-0.2cm}}
\name{Srinivas~Parthasarathy and~Carlos~Busso \vspace{-0.2cm} \thanks{This work was funded by NSF CAREER award IIS-1453781.}}
\address{Multimodal Signal Processing(MSP) lab, Department of Electrical and Computer Engineering\\
The University of Texas at Dallas, Richardson TX 75080, USA}
\email{sxp120931@utdallas.edu, busso@utdallas.edu \vspace{-0.0cm}}

\begin{document}
\maketitle
\begin{abstract}
Recognizing emotions using few attribute dimensions such as arousal, valence and dominance provides the flexibility to effectively represent complex range of emotional behaviors. Conventional methods to learn these emotional descriptors primarily focus on separate models to recognize each of these attributes. Recent work has shown that learning these attributes together regularizes the models, leading to better feature representations. This study explores new forms of regularization by adding unsupervised auxiliary tasks to reconstruct hidden layer representations. This auxiliary task requires the denoising of hidden representations at every layer of an auto-encoder. The framework relies on \emph{ladder networks} that utilize skip connections between encoder and decoder layers to learn powerful representations of emotional dimensions. The results show that ladder networks improve the performance of the system compared to baselines that individually learn each attribute, and conventional denoising autoencoders. Furthermore, the unsupervised auxiliary tasks have promising potential to be used in a semi-supervised setting, where few labeled sentences are available.
\end{abstract}
\noindent\textbf{Index Terms}: speech emotion recognition, regularization.

\vspace{-0.1cm}
\section{Introduction}
\label{sec:intro}
\vspace{-0.1cm}

Affective computing plays an important role in \emph{human computer interaction} (HCI). Emotions are conventionally represented with discrete classes such as happiness, sadness, and anger \cite{Busso_2004, Lee_2004, Schuller_2007}. An alternative emotional representation is through attribute dimensions such as arousal (calm versus active), valence (negative versus positive) and dominance (weak versus strong) \cite{Nicolaou_2011, Russell_1980, Fontaine_2007,Abdelwahab_2018}. These attribute dimensions provide the flexibility to represent multiple complex emotional behaviors, which cannot be easily captured with categorical descriptors. Furthermore, attribute dimensions can represent varying intensities of emotional externalizations which are lost when we use broad categorical descriptors such as ``anger'' (\eg cold versus hot anger). Therefore, constructing models that can accurately predict attribute scores is an important research problem.

Conventionally emotional attributes are individually modeled \cite{Wollmer_2008}, assuming that the attribute dimensions are orthogonal to each other. However, previous studies have shown significant correlation between different attributes \cite{Lewis_2007}. This observation strongly suggests the need for jointly modeling multiple emotional attributes. An appealing way to do this task is through \emph{multi-task learning} (MTL) where auxiliary tasks representing various emotional attributes are jointly learned \cite{Parthasarathy_2017_3,Chen_2017}. Learning the auxiliary task along with the primary task regularizes the learning process and the models generalize better.

While MTL generalizes the models, it requires labeled data (supervised auxiliary tasks). Regularization can also be performed with the help of unsupervised auxiliary tasks \cite{Hinton_2006_2, Ranzato_2008}. One appealing approach is the reconstruction of intermediate feature representations using autoencoders \cite{Hinton_2006_1}. Generally these unsupervised auxiliary tasks are performed as pre-training which is followed by normal supervised training of the primary task \cite{Hinton_2006_2}. The main criticism of this approach is that the feature representation learned by the autoencoder does not necessarily support the supervised classification or regression tasks, which require the learning of invariant features that are discriminative for the task.

This paper proposes ladder networks for emotion recognition, showing its benefits for emotional attribute predictions. Ladder networks conveniently solve unsupervised auxiliary tasks along with supervised primary tasks \cite{Rasmus_2015_1,Valpola_2015}. The unsupervised tasks (with respect to the primary task of predicting emotional attribute value) involve the reconstruction of hidden representations of a denoising autoencoder with lateral (skip) connections between the encoder and decoder layers. The representations from the encoder are simultaneously used to solve the supervised learning problem. The reconstruction of the hidden representations regularizes our primary regression task of predicting emotional attributes. The skip connections between the encoder and decoder ease the pressure of transporting information needed to reconstruct the representations to the top layers. Therefore, top layers can learn features that are useful for the supervised task, such as the prediction of emotional attributes. Interestingly, the framework also allows us to add multiple supervised tasks creating ladder networks with MTL structures.

This paper analyses the benefits of unsupervised auxiliary tasks to predict emotional attributes with ladder networks. We compare performance of these architectures with three baselines. The first baseline uses features from a denoising autoencoder that does not consider emotional labels to create the feature representation (\ie unsupervised autoencoder). The second baseline is the conventional supervised \emph{single task learning} (STL), where the emotional attributes are individually predicted. The third baseline is the MTL framework proposed by Parthasarathy and Busso \cite{Parthasarathy_2017_3}, which does not use unsupervised auxiliary tasks (\ie ladder networks). The performance shows that the architectures that use unsupervised auxiliary tasks consistently outperform the baselines. Furthermore, ladder networks with MTL structures have the best performance, improving the predictions of emotional attributes in the MSP-Podcast dataset.


\begin{figure*}[tb]
	\centering
	\subfigure[Autoencoder]
	{
		\includegraphics[width=2cm]{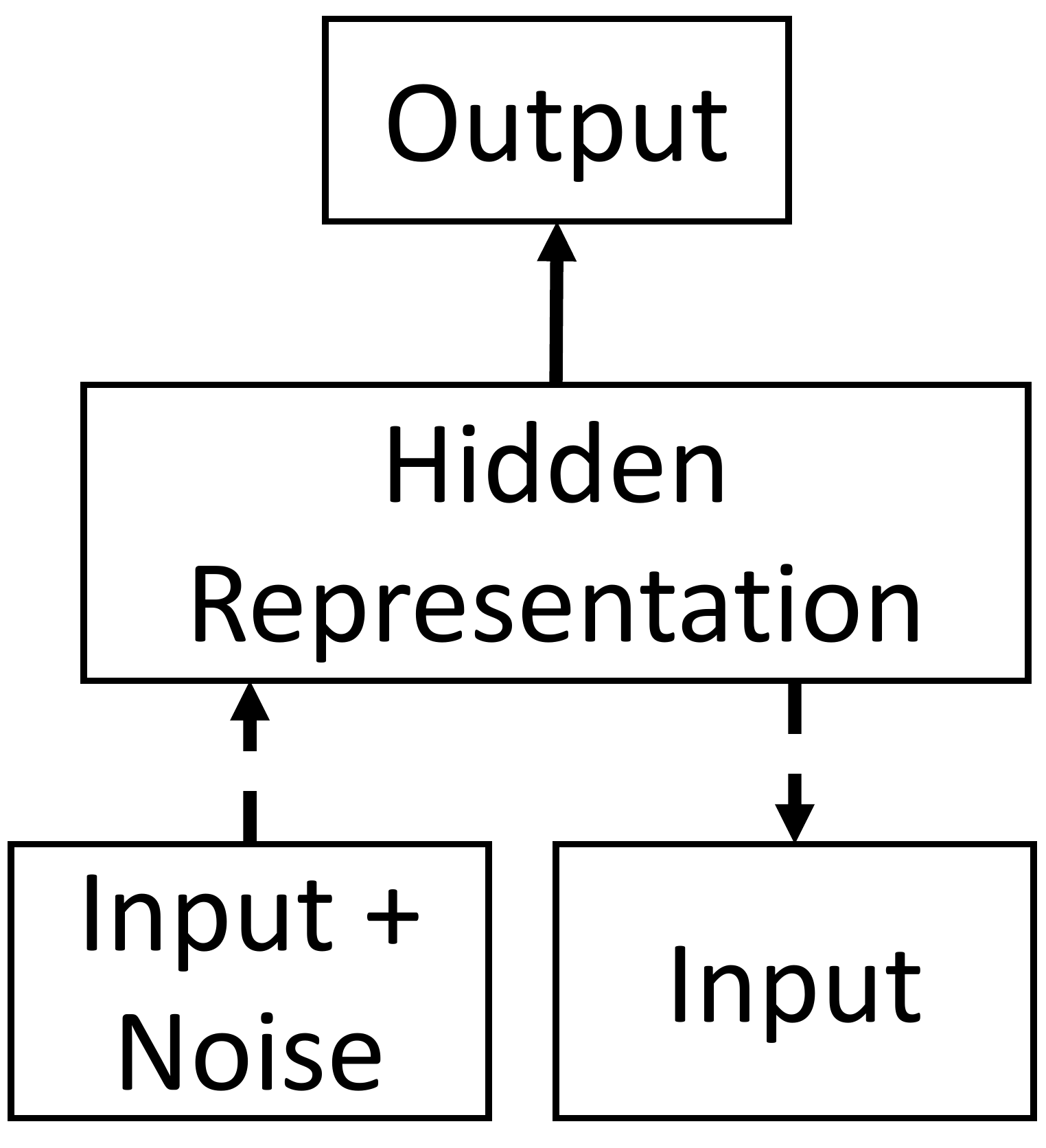}
		\label{fig:method_autoencoder}
	}\hspace{0.4cm}
	\subfigure[STL]
	{
		\includegraphics[width=2cm]{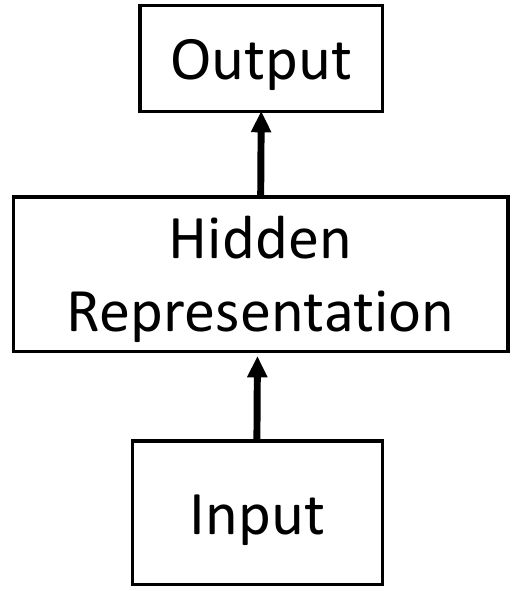}
		\label{fig:method_stl}
	}\hspace{0.4cm}
	\subfigure[MTL]
	{
		\includegraphics[width=3.5cm]{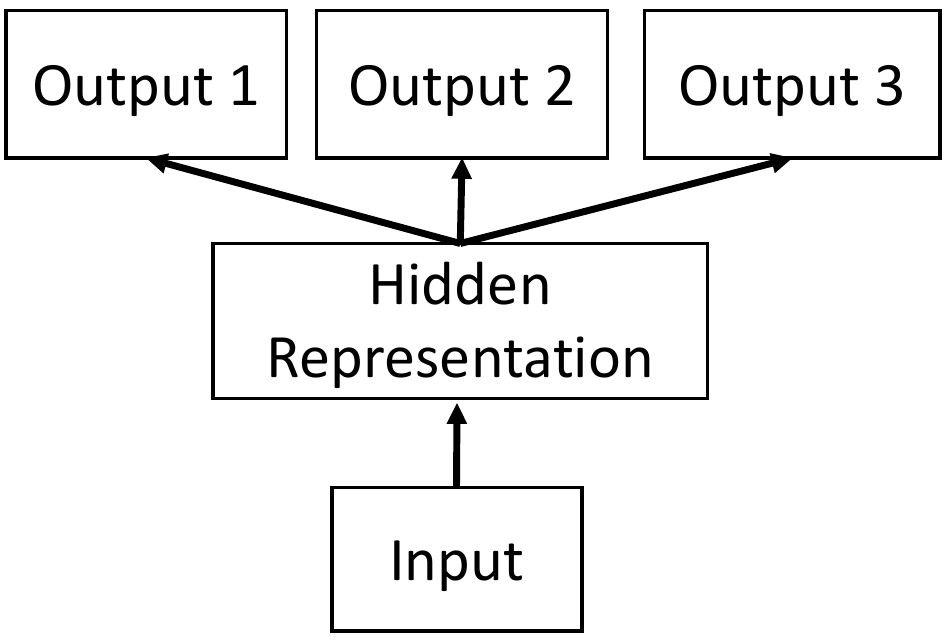}
		\label{fig:method_mtl}
	}\hspace{0.4cm}
	\subfigure[Ladder-STL]
	{
		\includegraphics[width=2cm]{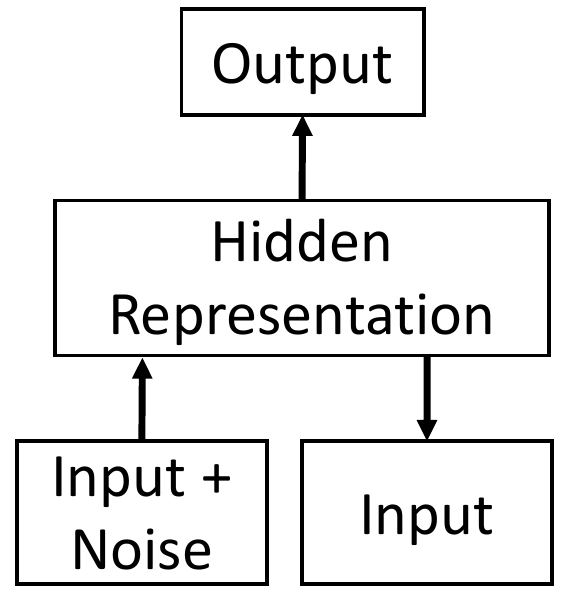}
		\label{fig:method_ladder_stl}
	}\hspace{0.4cm}
	\subfigure[Ladder-MTL]
	{
		\includegraphics[width=3.5cm]{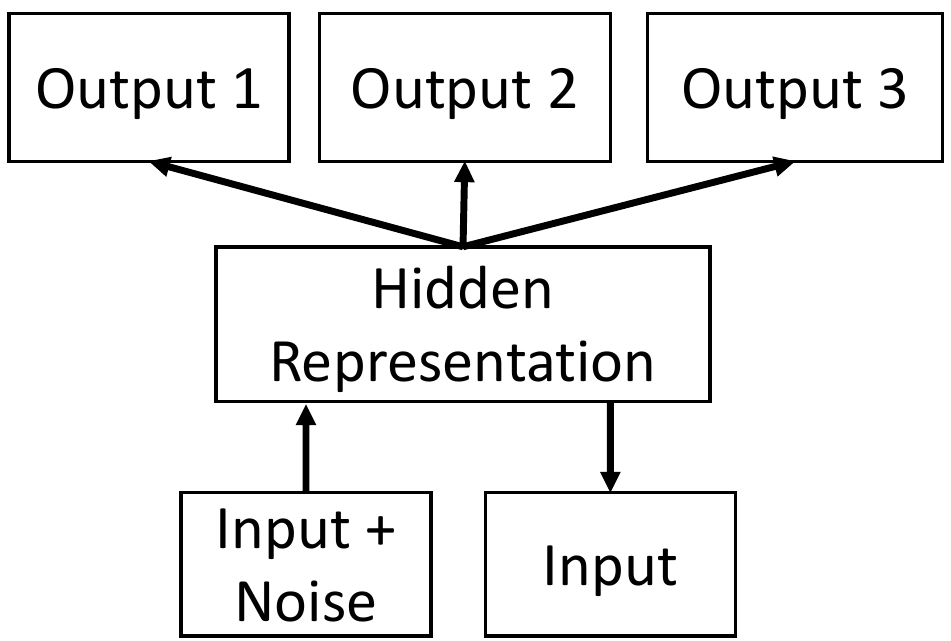}
		\label{fig:method_ladder_mtl}
	}
    \vspace{-0.3cm}
	\caption{Various architectures with supervised and unsupervised auxiliary tasks for emotion prediction. The first three structures are the baseline systems. The last two are the proposed models with ladder networks.}
	\label{fig:methodology}
    \vspace{-0.3cm}
\end{figure*}

\vspace{-0.3cm}
\section{Background}
\label{sec:background}
\vspace{-0.15cm}

\subsection{Related Work}
\label{ssec:related}
\vspace{-0.15cm}

Few studies have focused on using auxiliary tasks to improve emotion recognition. Parthasarathy and Busso \cite{Parthasarathy_2017_3} proposed the joint learning of arousal, valence, dominance through \emph{multi-task learning} (MTL). They showed significant improvement in performance when attributes are jointly predicted compared to single task learning. Chen \etal \cite{Chen_2017} jointly learned arousal and valence for continuous emotion recognition, leveraging the relationship between the attributes. Their approach achieved improved performance for the affect subtask on the AVEC 2017 challenge. Xia and Liu \cite{Xia_2015} proposed a scheme to use the regression of emotional attributes as auxiliary task to aid the classification of emotional classes. Chang and Scherer \cite{Chang_2017} proposed a valence classifier that used predictions on arousal as a secondary task. Their MTL framework did not show any improvement over learning just the primary task. Kim \etal \cite{Kim_2017} proposed using gender and naturalness of data as auxiliary tasks to recognize emotions. Finally, Le \etal \cite{Le_2017} proposed a classifier that continuously recognized emotional attributes. The attribute values were discretized using the $k$-means algorithm with $k\in \{4, 6, 8, 10\}$. The discretized attribute values were treated as classes which were jointly predicted with MTL.

The proposed approach builds upon ladder networks that effectively combine supervised classification or regression problems with unsupervised auxiliary tasks. Valpola \cite{Valpola_2015} proposed using lateral shortcut connections to aid deep unsupervised learning. Rasmus \etal \cite{Rasmus_2015_1,Rasmus_2015_2} further extended this idea to support supervised learning. They included a batch normalization to reduce covariate shift. They also compared various denoising functions to be used by the decoder. Finally, Pezeshki \etal \cite{Pezeshki_2016} studied the various components that affected the ladder network, noting that lateral connections between encoder and decoder and the addition of noise at every layer of the network greatly contributed to their improved performance. We describe in detail this framework in Section \ref{ssec:ladder}.

\vspace{-0.2cm}
\subsection{Database}
\label{ssec:database}
\vspace{-0.1cm}

This study uses the version 1.1. of the MSP-Podcast dataset \cite{Lotfian_201x}. The dataset contains emotionally colored, naturalistic speech from podcasts downloaded from audio sharing websites. The podcasts are processed and further split into smaller segments between 2.75s and 11s of duration. The segments are long enough so meaningful features can be extracted, and short enough so the emotional content does not change across the speaking turn. The dataset contains 22,630 (38h56m) audio segments. We manually identified the speaker identity of 18,991 sentences spoken by 265 speakers. The speaker information is used to create the train, development and test partitions. The partitions aim to have speaker independent sets. The test set contains 7,181 segments from 50 speakers, the development set contains 2,614 sentences from 20 speakers, and the train set includes the rest of the corpus (12,835 sentences). Audio segments are emotionally annotated on \emph{Amazon Mechanical Turk} (AMT). The annotations are collected through perceptual evaluations from five or more raters. 
The emotional attributes are annotated with \emph{self-assessment manikins} (SAMs) on a seven likert-scale for arousal (1-very calm, 7-very active), valence (1- very negative, 7-very positive), and dominance (1- very weak, 7 - very weak). The ground-truth labels for the attributes of a segment are the average scores provided by the evaluators.

\vspace{-0.2cm}
\section{Methodology}
\label{sec:methodology}
\vspace{-0.1cm}

\subsection{Proposed method}
\label{ssec:proposed}
\vspace{-0.1cm}

An important challenge while designing emotion recognition systems is to make models that generalize across different conditions \cite{Busso_2013}. Conventional models (Figure \ref{fig:method_stl}) show poor performance when trained and tested on different corpora \cite{Shami_2007,Parthasarathy_2017_3}. Therefore, regularizing deep learning models is crucial in emotion recognition to find representations that are not overfitted to a particular domain. Regularization can be implemented with various approaches including early stopping criterion and dropout. The approach proposed is this study is to solve auxiliary tasks along with the primary task of predicting emotional attributes. By training models that are optimized for primary and auxiliary tasks, the feature representations are more general, avoiding overfitting. 
There are multiple ways to introduce auxiliary tasks to model emotion recognition. Previous studies for emotion recognition have focused on supervised auxiliary tasks, involving learning multiple emotion attributes \cite{Parthasarathy_2017_3} (Figure \ref{fig:method_mtl}), combining emotional classification problem with regression of emotional attributes \cite{Xia_2015}, and learning other labels such as gender and age along with the emotion \cite{Kim_2017}. While these approaches are appealing, the supervised nature of the tasks require auxiliary labels for the training samples. Labels for emotional data commonly come from perceptual evaluations where multiple raters judge the emotional content of the stimuli. These evaluations are both expensive and time consuming. Therefore, annotating additional meta-information is not a feasible alternative. Its appealing to create unsupervised auxiliary tasks to regularize the network. 

\begin{figure*}[tb]
	\centering
	\subfigure[Ladder Network]
	{
		\includegraphics[width=12cm]{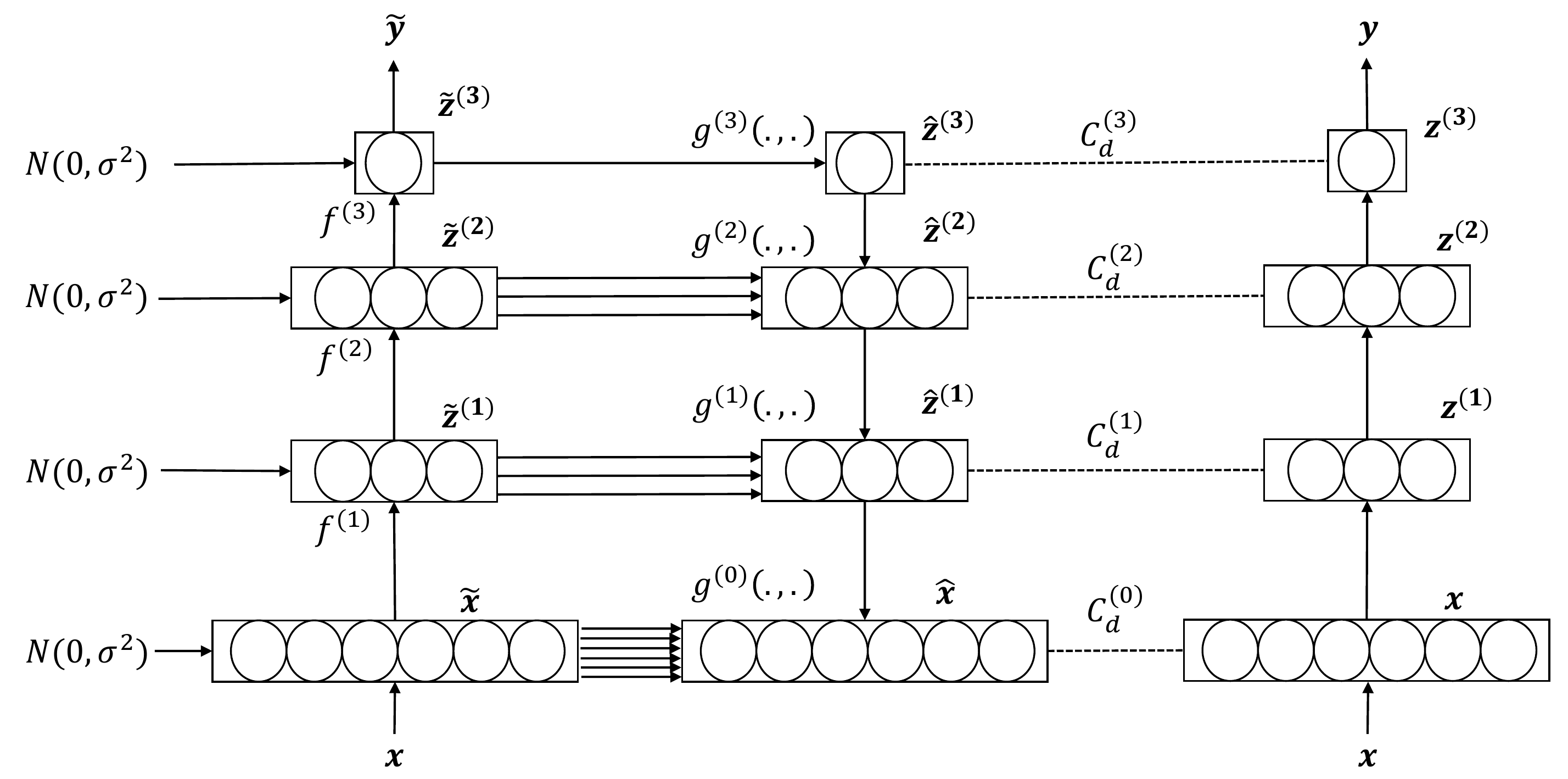}
		\label{fig:ladder}
	}
	\subfigure[MTL Network]
	{
		\includegraphics[width=2.77cm]{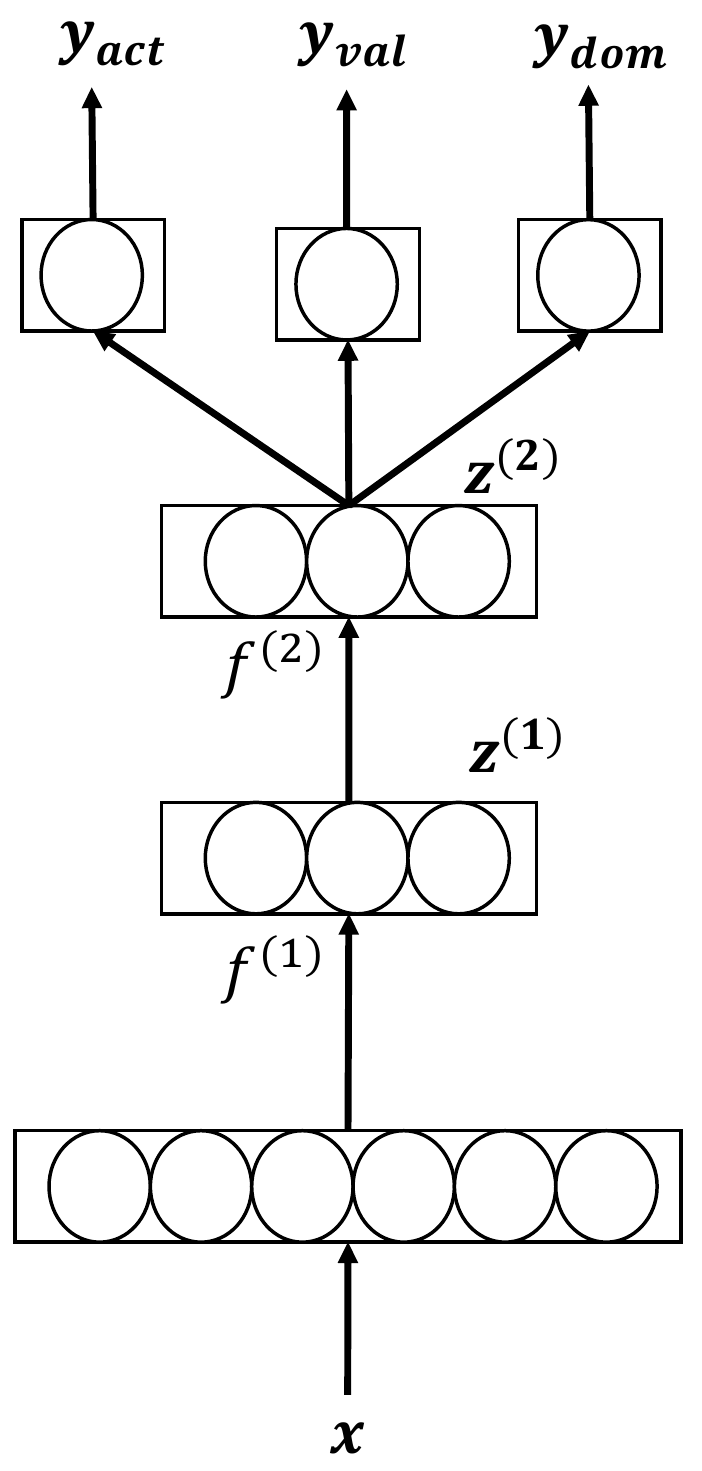}
		\label{fig:mtl}
	}
    \vspace{-0.3cm}
	\caption{Architectures using auxiliary tasks for emotion attribute prediction. \ref{fig:ladder} illustrates the ladder network with unsupervised auxiliary tasks. \ref{fig:mtl} illustrates the MTL network that jointly learns multiple attribute values.}
    \vspace{-0.2cm}
	\label{fig:structure}
\end{figure*}

While traditional autoencoders (Figure. \ref{fig:method_autoencoder}) reconstruct input features in an unsupervised fashion, the intermediate latent representations are not trained for the underlying regression or classification task. This paper proposes to employ the unsupervised reconstruction of inputs as an auxiliary task to regularize the network, while optimizing the performance of an emotion regression system. We efficiently achieve this goal with ladder network architectures (Figure. \ref{fig:method_ladder_stl}). The addition of an unsupervised auxiliary task not only regularizes the learning of the primary task, but also helps learning powerful discriminative representations of the input features. Furthermore, since there is no constraint on the primary task itself, we can combine it with other supervised auxiliary tasks to produce powerful models for predicting emotional attributes (Figure. \ref{fig:method_ladder_mtl}).



\vspace{-0.2cm}
\subsection{Ladder Networks}
\label{ssec:ladder}
\vspace{-0.1cm}

Ladder networks combine supervised primary task with unsupervised auxiliary tasks. The auxiliary tasks reconstruct the hidden representations of a denoising autoencoder. The encoder of the autoencoder is simultaneously used to train the primary task at hand. The key aspect of the ladder network is the lateral connections between encoder and decoder layers. These skip connections allow the decoder to directly learn the representation from the encoder layer, bypassing the top layers of the encoder which can then learn representations that would help with the primary supervised task. Figure. \ref{fig:ladder} illustrates a conceptual ladder network with two hidden layers used for a regression task. Note that the true benefits of the ladder network is for semi-supervised setting where few labeled samples for the primary task are available in the target domain. However, this study focuses on the fully-supervised setting where we have emotional labels for every sample. The semi-supervised case is left as a future work.

\noindent\textbf{Encoder:} The encoder of the ladder network is a fully connected \emph{multilayer perceptron} (MLP) network. 
A Gaussian noise with variance $\sigma^2$ is added to each layer of the noisy encoder (Figure. \ref{fig:ladder}). The representation from the final layer $\tilde{\mathbf{z}}^{(L)}$ of the encoder is used as the target for the supervised task. The decoder tries to reconstruct the latent representation $\hat{\mathbf{z}}$ at every layer using, as target, a clean copy of the encoder \textbf{z}.
Note that the supervised task, in this study the prediction of emotional attributes, is trained with the noisy encoder which further regularizes the supervised learning. However, for inference we use the predictions from the clean encoder. We describe the choice of hyper-parameters for the network in Section \ref{ssec:parameters}.

\noindent\textbf{Decoder:} 
The goal of the decoder is to denoise the noisy latent representations. The denoising function, $g()$, in Figure \ref{fig:ladder} combines top-down information from the decoder and the lateral connection from the corresponding encoder layer. With lateral connections, the ladder networks perform similar to hierarchical latent variable models. Lower layers are mostly responsible for reconstructing the input vector. This approach allows higher layers to learn more abstract, discriminative features needed for the supervised task.
We use the denoising function proposed by Pezeshki \etal \cite{Pezeshki_2016}, modeled by an MLP with inputs $[u, \tilde{z}, u \odot z]$, where $u$ is the batch normalized projection of the layer above and $\odot$ represents the Hadamard product. We use an MLP with 1 hidden layer and 4 hidden nodes to model the denoising function $g()$. The overall loss function is:



\begin{table*}[tb]		
	\caption{CCC values for the validation and test sets. The evaluations include 256 nodes per layer for arousal (\emph{Aro}), valence (\emph{Val}), and dominance (\emph{Dom}). \textbf{Bold values} indicate the model(s) with the best performance per attribute (multiple cases are highlighted when differences are statistical significant). $^\ast$ indicates significant improvements of the ladder networks compared to all the baselines.}
	\centering
	\begin{tabular*}{2\columnwidth}{@{\extracolsep{\fill}}c|c|c|c||c|c|c}
		\hline
		\multirow{2}{*}{Task}& \multicolumn{3}{c||}{Validation}& \multicolumn{3}{c}{Test}\\
		\cline{2-7}
		\cline{2-7}
		& Aro & Val & Dom & Aro & Val & Dom\\
		\hline
		\hline
		Autoencoder & 0.358 $\pm$ 0.069  & 0.136 $\pm$ 0.141	& 0.305 $\pm$ 0.139 & 0.272 $\pm$ 0.136 & -0.006 $\pm$ 0.012 & 0.284 $\pm$ 0.148  \\
		STL & 0.778 $\pm$ 0.004	& 0.443 $\pm$ 0.008 & 0.722 $\pm$ 0.004
		& 0.737 $\pm$ 0.008 & \textbf{0.292 $\pm$ 0.007}  & 0.670 $\pm$ 0.007\\
		MTL & 0.791 $\pm$ 0.003  & \textbf{0.469 $\pm$ 0.010} & 0.735 $\pm$ 0.003 & 0.745 $\pm$ 0.008 & 0.285 $\pm$ 0.007 & 0.676 $\pm$ 0.006\\
        \hline
		Ladder+STL & 0.801 $\pm$ 0.002$^\ast$ & 0.443 $\pm$ 0.007 & 0.742 $\pm$ 0.002$^\ast$ & \textbf{0.765 $\pm$ 0.002}$^\ast$ & \textbf{0.294 $\pm$ 0.007} & 0.687 $\pm$ 0.003$^\ast$\\
		Ladder+MTL & \textbf{0.803 $\pm$ 0.002}$^\ast$ & 0.458 $\pm$ 0.004 & \textbf{0.746 $\pm$ 0.001}$^\ast$ & 0.761 $\pm$ 0.002$^\ast$ & \textbf{0.289 $\pm$ 0.008} & \textbf{0.689 $\pm$ 0.002}$^\ast$\\
		\hline
	\end{tabular*}
	\vspace{-0.1cm}
	\label{tab:results}
\end{table*}

\begin{align}
C = C_c + \lambda_l \sum_l C_d^{(l)}
\label{eq:CostLadder}
\end{align}

\noindent where $C_c$ is the supervised loss, $C_d^{(l)}$ is the reconstruction loss at layer $l$ and $\lambda_l$ is a hyper-parameter weight for the loss. 

\vspace{-0.2cm}
\subsection{Ladder Network with Multi-task learning}
\label{ssec:mtl}
\vspace{-0.1cm}

While ladder network utilizes the reconstruction cost as an unsupervised auxiliary task, regularization can also be achieved through supervised tasks. With emotional attributes, MTL can achieve this goal by joint learning multiple attributes as proposed by Parthasarathy and Busso \cite{Parthasarathy_2017_3}. Figure \ref{fig:mtl} illustrates a two hidden layer MTL network that jointly predicts three emotional attributes: arousal, valence and dominance. The overall loss for the MTL network is given by

\vspace{-0.2cm}

\begin{align}
C_{\mathit{MTL}} = \alpha C_{\mathit{aro}} + \beta C_{\mathit{val}} + (1 - \alpha - \beta) C_{\mathit{dom}}
\label{eq:costMTL}
\end{align}

\noindent
with $0 < \alpha, \beta < 1$ and $\alpha + \beta \leq 1$. Parthasarathy and Busso \cite{Parthasarathy_2017_3} showed that MTL networks perform better than STL networks for predicting emotional attributes. 
An appealing framework is to combine the proposed ladder network with MTL as shown in Figure \ref{fig:method_ladder_mtl}. The supervised loss $C_c$ in Equation \ref{eq:CostLadder} is replaced by the MTL loss $C_{\mathit{MTL}}$ from Equation \ref{eq:costMTL}. This approach aims to combine supervised and unsupervised auxiliary tasks, creating an architecture that can learn powerful feature representations targeted for the prediction of emotional attributes. 



\vspace{-0.2cm}
\section{Experimental Evaluation}
\label{sec:experiments}
\vspace{-0.1cm}
\subsection{Acoustic Features}
\label{ssec:feature}
\vspace{-0.1cm}

We use the feature set introduced for the Computational Paralinguistic Challenge at Interspeech 2013 \cite{Schuller_2013}. The feature extraction process involves two parts. First, \emph{low level descriptors} (LLDs) are extracted on a frame-by-frame basis. The set includes \emph{mel frequency cepstral coefficients} (MFCCs), fundamental frequency (F0) and energy. Various statistics, denoted as \emph{high level features} (HLFs) are calculated over the LLDs. Overall, the feature set contains 6,373 features which we use for the various tasks in this study. The features were extracted with OpenSMILE \cite{Eyben_2010_2}.

\vspace{-0.2cm}
\subsection{Baselines}
\label{ssec:baseline}
\vspace{-0.1cm}

We compare our results with three baseline networks. Figure \ref{fig:method_stl} shows the first baseline, which is a \emph{deep neural network} (DNN) separately trained for each emotional attribute (i.e., STL). Figure \ref{fig:method_autoencoder} shows the second baseline, which learns feature representations using an autoencoder in an unsupervised fashion. The feature representations learned are then used as input for the supervised task. Unlike the ladder networks, the feature representation is independently learned from the supervised task. The objective of the autoencoder is to learn hidden representations that are useful for denoising the noise added to the features. All weights and activations of the encoder are frozen, and an output layer is then added on the top layer of the encoder for the prediction task. 
Figure \ref{fig:method_mtl} shows the third baseline, which uses MTL to jointly predict the three emotional attributes. Following the work of Parthasarathy and Busso \cite{Parthasarathy_2017_3}, we train three MTL networks, one for each target emotional attribute, optimizing $\alpha$ and $\beta$ in Equation \ref{eq:costMTL} to maximize the performance for each attribute. By learning all three tasks, we obtain feature representations that generalize well across different conditions. 


\vspace{-0.2cm}
\subsection{Implementation Details}
\label{ssec:parameters}
\vspace{-0.1cm}

All the deep neural networks in this study have two hidden layers with 256 nodes per layer. We use \emph{rectified linear unit} (ReLU) activation at the hidden layers and a linear activation for the output layer. We optimize the networks using NADAM with a learning rate of $5e^{-5}$. The networks are implemented with dropout $p=0.5$ at the input and first hidden layer. 
Following Trigeorgis \etal \cite{Trigeorgis_2016}, we use the \emph{concordance correlation coefficient} (CCC) as the loss function for training the models. We also use CCC to evaluate the models. All the hyper-parameters are set maximizing performance on the validation set, including the parameters for MTL ($\alpha, \beta$). We train all the networks for 50 epochs with early stopping based on the results observed in the validation set. The best model on the validation set is then evaluated on the test set. All the models are trained 10 times with different random initializations, reporting the mean CCC.

For the ladder network, we add noise with variance $\sigma^2$=0.3 to each layer of the encoder. We conducted a grid search for the weights for the reconstruction loss with values $\lambda_l \in \{0.1, 1, 10 ,100\}$. A value of $\lambda_l=1$ gives the best result on the validation set. We use the mean squared error as the reconstruction cost and a dropout with probability $p=0.1$ at the input layer and the first hidden layer. Notice that the original paper implementing ladder network did not use dropout. However, the high dimensionality of our feature vector and the violation of the independence assumption in our features motivate us to further regularize the network by adding dropout.

\vspace{-0.2cm}
\section{Results}
\label{sec:results}

Table \ref{tab:results} illustrates the mean CCC and standard deviation of the proposed architectures and the baselines for the validation and test sets. We analyze the performance of various models using the one-tailed $t$-test over the 10 trials, asserting significance if $p$-value$<$0.05. We highlight with an asterisk when the ladder models perform better than the baselines.
First, we note that the results on the validation set are significantly higher than performance on the test set for all emotional attributes and all models. Comparing the performance of the different architectures on the test set, we note that all the networks perform better than the autoencoder baseline. This result shows that the feature representation learned by the autoencoder does not fit well for the primary regression task. Next, amongst the baselines we see that MTL models perform the best in all cases, except for valence on the test set. This result further confirms the benefits of supervised auxiliary tasks through the joint learning of multiple emotional attributes, shown in our previous study \cite{Parthasarathy_2017_3}. Observing the proposed architectures, we note that in almost all cases the ladder networks perform significantly better than the baselines and give the best performance. For valence, while MTL performs better than all other methods on the validation set, it does not translate to the test set where the Ladder+STL architecture has the best performance. The role of regularization for valence is an interesting topic which requires further study. Amongst the proposed architectures, Ladder+MTL performs better than Ladder+STL in many cases. The results show the benefits of combining both unsupervised and supervised auxiliary tasks for predicting emotional attributes. Overall, the unsupervised auxiliary tasks greatly help to regularize our network improving the predictions and providing yet state-of-the-art performance on the MSP-Podcast corpus.

\vspace{-0.2cm}
\section{Conclusions}
\label{sec:conclusion}
\vspace{-0.1cm}

This work proposed ladder networks with multi-task learning to predict emotional attributes, achieving state-of-the-art performance on the MSP-Podcast corpus. We illustrated the benefits of using auxiliary tasks to regularize the network by combining unsupervised auxiliary tasks (ladder network) and supervised auxiliary tasks (multi-task learning). The emotional models generalize better across various conditions providing significantly better performance than the baseline systems.

There are many future directions for this work. First, the true potential of using unsupervised auxiliary tasks is in harnessing the almost unlimited amount of unlabeled data. We can extend the framework to work in a semi-supervised manner, where we can combine large number of unlabeled samples with fewer emotionally labeled samples, providing a more powerful feature representation. Second, the feature representations produced by the ladder network can be further studied and used as general input features for other emotion recognition problems such as classification of emotional categories. Finally, the auxiliary tasks could be extended to cover multiple modalities generalizing the models even more. The promising results in this study suggest that these extensions can lead to important improvements in emotion prediction performance. 

\balance

\bibliographystyle{IEEEtran}
\bibliography{reference}


\end{document}